\documentclass[prl, preprint, superscriptaddress, showpacs]{revtex4}
\pdfoutput=1
\usepackage{graphicx}
\usepackage{amsmath}
\usepackage{hyperref}

\makeatletter
\def\btt#1{\texttt{\@backslashchar#1}}
\DeclareRobustCommand\bblash{\btt{\@backslashchar}}
\makeatother

\newcommand{\vek}[1]{\textbf{#1}}

\begin{document}
\bibliographystyle{apsrev}
\title{Implementation of non-equilibrium vertex corrections in KKR: transport through disordered layers}
\author{Christian Franz}
\affiliation{I. Physikalisches Institut, Justus Liebig University, Giessen, Germany}
\author{Michael Czerner}
\affiliation{I. Physikalisches Institut, Justus Liebig University, Giessen, Germany}
\author{Christian Heiliger}
\email{Christian.Heiliger@physik.uni-giessen.de}
\affiliation{I. Physikalisches Institut, Justus Liebig University, Giessen, Germany}

\date{\today}

\begin{abstract}
The theoretical description of modern nanoelectronic devices requires a quantum mechanical treatment and often involves
disorder, e.g.\ form alloys. Therefore, the \textit{ab initio} theory of transport using non-equilibrium Green's
functions is extended to the case of disorder described by the coherent potential approximation. This requires the
calculation of non-equilibrium vertex corrections. We implement the vertex corrections in a Korringa-Kohn-Rostoker
multiple scattering scheme. In order to verify our implementation and to demonstrate the accuracy and applicability we
investigate a system of an iron-cobalt alloy layer embedded in copper. The results obtained with the coherent potential
approximation are compared to supercell calculations. It turns out that vertex corrections play an important role for
this system.
\end{abstract}
\maketitle

\subsection{Introduction}

Modern applications require an accurate description of transport processes on the nanometer scale, where quantum
coherence effects cannot be neglected and thus fully quantum mechanical calculations are necessary. A popular tool for
these calculations is the non-equilibrium Green's function (NEGF) method (or Keldysh NEGF)~\cite{datta1999}. This method
can be implemented within the Korringa-Kohn-Rostoker (KKR) multiple scattering scheme~\cite{henk2006,heiliger2008}.
Advantages of the NEGF method include the capability to calculate the transport under applied bias
self-consistently~\cite{brandbyge2002,rocha2006}, the simple inclusion of inelastic scattering events by additional
self-energies~\cite{datta1999}, and a numerically more stable description of the transmission coefficient in comparison
to methods using the current operator~\cite{henk2006}.

The presence of chemical disorder from alloying breaks the translational symmetry of the crystal and thus complicates the
calculation. In order to restore the periodicity one can use an effective medium scheme, i.e.\ the alloy is replaced by
an effective medium which approximates the properties of the alloy but has the periodicity of the underlying lattice. The
probably most popular one is the coherent potential approximation (CPA)~\cite{zabloudil}. Here, the effective medium is
defined by a self-consistent condition. The idea is that the effective medium Green's function (GF) should resemble the
full configurational average of the alloy GF. Therefore, the single-site CPA condition requires that the additional
scattering from a real atom placed on one site in the effective medium crystal should average to zero when the alloy
average is taken. While this approximation is very successful in many applications, it fails to take into account the
effect of correlations due to local clusters in the alloy. The single-site CPA can be extended to include several sites,
e.g.\ in the nonlocal CPA~\cite{rowlands2003}, thus taking the effects of short range order into account.

An alternative approach is the use of large supercells averaging over a large number of configurations. The drawbacks of
this method are high computational costs, possible supercell effects, and the limited number of realizable
concentrations. It has the advantage that it can be applied within most band structure methods without additional
implementations. Further, the accuracy of the description can be enhanced simply by using larger supercells and more
configurations. In this sense, it is possible, by convergence test with respect to the size of the supercell, to find the
results of the real system. In practice, this is rarely possible due to the high computational effort, in particular, for
transport calculations. However, for simple systems the supercell approach can be used to test the validity of the CPA
results.
In this contribution we focus on chemical disorder. Nevertheless, the CPA can be used to describe other kinds of
disorder, e.g.\ thermal disorder like lattice vibrations or magnetic fluctuations~\cite{butler1985,ebert2011}.

Performing transport calculations combined with CPA leads to wrong results if one just uses the transport formulas with
the effective medium GF. In a simple picture, the transport formulas contain products of two GFs. Using only the
effective medium GF we have products of averaged GFs, which is not the same as the average of the product of GFs. From a
physical perspective, since the effective medium GF only includes the damping of Bloch waves due to disorder, this
approach would drop electrons scattered by disorder instead of scattering them to a different state. Therefore, so
called vertex corrections have to be taken into account, which describe the scattered electrons. These were derived for the case when
the operator between the GFs does not depend on the alloy configuration~\cite{velicky1969}. The case of a configuration
dependent operator like the current operator is more involved but leads to similar expressions~\cite{butler1985}. Within
the NEGF scheme the operator in question is the self-energy of a lead and does not depend on the alloy configuration. For
this case, the vertex corrections were derived for the linear muffin-tin orbital (LMTO) method~\cite{carva2006,ke2008}.
In this paper we present a detailed derivation of the NEGF implementation in KKR~\cite{heiliger2008} and report the, to
our knowledge, first implementation of vertex corrections within the KKR-NEGF-scheme. We test our implementation by
investigating transport in FeCo alloys comparing supercell and CPA results.

\subsection{Non-Equilibrium Green's Functions Method}

\begin{figure}
\centering
\includegraphics[width=0.7\textwidth]{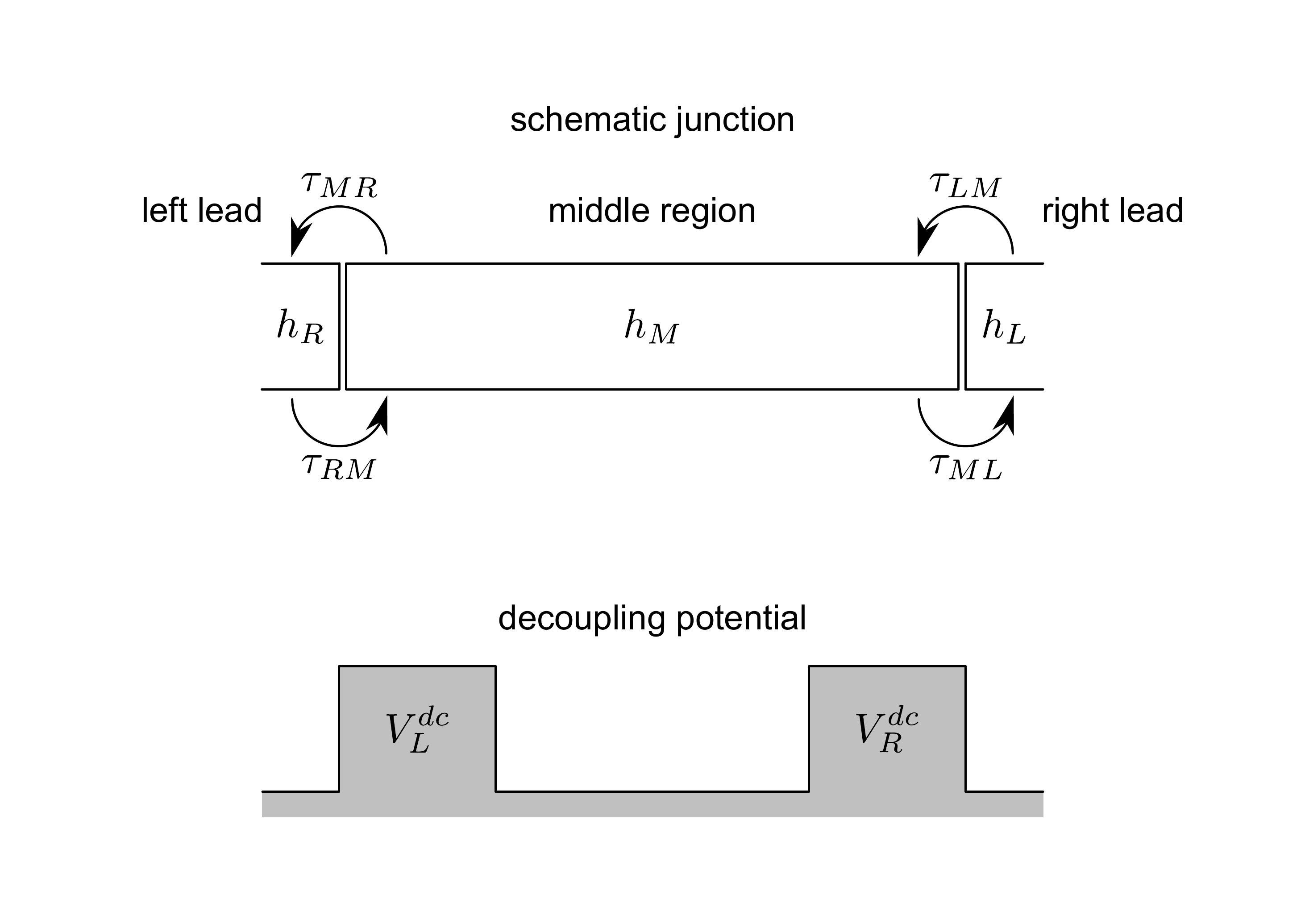}
\caption{Schematic picture of the junction and the decoupling potential.}
\label{geometry}
\end{figure}

The NEGF method was successfully implemented in the KKR scheme~\cite{heiliger2008,henk2006}. Here, we present a more
detailed and more general derivation of the used formulas. The usual way to apply the NEGF formalism to a system with
steady state transport is to divide the system of interest into three regions along the transport
direction~\cite{datta1999}: a middle region connected to a left and a right lead (see Fig.~\ref{geometry}). The leads are
considered as ideal conductors while the middle region contains the interesting part e.g.\ the alloy or a potential
barrier.
To describe the influence of the leads on the middle region one considers a fictitious decoupled
system with a Hamiltonian
\begin{equation}
H_{dc}=\left(\begin{array}{ccc}
    h_{L} & 0 & 0 \\
    0 & h_{M} &0 \\
    0 & 0 & h_{R}
\end{array}\right)
\end{equation}
and the corresponding Green's function (GF)
\begin{equation}
\tilde{G}_{dc}= \left[  z - H_ {dc} \right]^{-1} =
\left(\begin{array}{ccc}
    \tilde{G}_{L} & 0 & 0 \\
    0 & \tilde{G}_{M} &0 \\
    0 & 0 & \tilde{G}_{R}
\end{array}\right) .
\end{equation}
Here, $z=E + i\; \eta$ is the complex energy. We will always assume $\eta \geq 0$, thus $G(E)=\lim_{z \to E} G(z)$ is a
retarded and $G(E)^\dagger=\lim_{z \to E} G(z^*)$ an advanced GF. All physical quantities (density, transmission, etc.)
are obtained by taking the limit $z \to E$, i.e.\ $\eta \to 0^+$. In practice, this limit is taken by using a
sufficiently small $\eta$, determined by convergence tests. The coupling between the leads and the interstitial region
is given by the coupling matrix
\begin{equation}
\tau=\left(\begin{array}{ccc}
    0 & \tau_{LM} & 0 \\
     \tau_{ML} &0 &  \tau_{MR} \\
    0 & \tau_{RM}  & 0
\end{array}\right) .
\end{equation}
This includes the assumption that there is no direct coupling between the left and the right lead only indirect over the
middle region. Thus, the full system is described by the Hamiltonian $H_c=H_{dc}+\tau$ and the overall GF of the
coupled system
\begin{equation}
G_c=[z - H_{dc} - \tau]^{-1}=
\left(\begin{array}{ccc}
    G_{LL} & G_{LM} & G_{LR} \\
    G_{ML} & G_{MM} & G_{MR} \\
    G_{RL} & G_{RM} & G_{RR}
\end{array}\right) .
\end{equation}
By solving this system of equations for $G_{MM}$, the GF of the middle region coupled to the leads, it can be described
by
\begin{equation}
G_{MM}=\left[ z - h_M - \Sigma \right]^{-1} ,
\end{equation}
where the so called self-energy of the leads is introduced
\begin{eqnarray}
\Sigma     &=& \Sigma_{L} + \Sigma_{R} \label{SE1},\\
\Sigma_{L} &=& \tau_{ML} \; \tilde{G}_L \; \tau_{LM} \nonumber,\\
\Sigma_{R} &=& \tau_{MR} \; \tilde{G}_R \; \tau_{RM} \nonumber.
\end{eqnarray}
These self-energies can be interpreted as fluxes of incoming and outgoing electrons at the connection between leads and
middle region~\cite{datta1999}. They describe the influence of the semi-infinite leads on the middle region and are
determined by the surface GFs of the leads and the coupling to the middle region. The coupled and decoupled GFs in the
middle region are connected by the self-energies via the Dyson equation
\begin{equation}\label{SE2}
G_{MM} = \tilde{G}_{M} + \tilde{G}_{M} \; \Sigma_L \; G_{MM} + \tilde{G}_{M} \; \Sigma_R \; G_{MM} .
\end{equation}

Using the self-energies and the coupled GF the density can be calculated~\cite{kadanoff1962,datta1999,haug2008}
\begin{equation}\label{nequ_den}
  n(E) = \frac{i}{\pi} G_{MM}(E) \; \Gamma(E) \; G_{MM}(E)^\dagger ,
\end{equation}
where $\Gamma(E) = i\;(\Sigma(E)-\Sigma(E)^\dagger)$. The density can be decomposed into the share of electrons
originating from the left or right lead by using $\Gamma_L=i\;(\Sigma_L-\Sigma_L^\dagger)$ or
$\Gamma_R=i\;(\Sigma_R-\Sigma_R^\dagger)$ instead of $\Gamma = \Gamma_L + \Gamma_R$.
This way, it is possible to obtain the non-equilibrium density or in other words the density of the transport electrons.
This is for example required to calculate the spin-transfer torque~\cite{heiliger2008,heiliger20081}. Another
application is the analysis of the transport states e.g.\ during tunneling.

On the other hand, the equilibrium density can also be calculated by~\cite{zabloudil}
\begin{equation}\label{equ_den}
  n(E) = -\frac{1}{\pi} \mathrm{Im} \left[ G_{MM}(E) \right] .
\end{equation}
Now, $n(E)$ in Eq.~(\ref{nequ_den}) involves a product of GFs, whereas Eq.~(\ref{equ_den}) does not. Therefore, for
$n(E)$ in Eq.~(\ref{nequ_den}) vertex corrections are required in the presence of CPA. But both have to give the same
results in equilibrium. Consequently, this provides a simple and stringent way to test the vertex corrections, including
their convergence.

Many transport properties of the system can be calculated from the transmission function $T(E)$, which is given
by~\cite{fisher1981,datta1999}
\begin{equation}\label{TE}
T(E) = \mathrm{Tr} \Big[ G_{MM}(E) \; \Gamma_L(E) \; G_{MM}(E)^\dagger  \;\Gamma_R(E)
\Big] ,
\end{equation}
where the trace is over the basis set.

In the KKR method, the GF of the coupled system is known and can be calculated via the decimation
technique~\cite{zabloudil}. A difficulty occurs when describing the self-energy in terms of the decoupled system. The KKR
method works with GFs and has no finite couplings $\tau$ which could be set to zero. To solve this problem, one can
introduce an artificial potential barrier $V^{dc}=V^{dc}_{L}+V^{dc}_{R}$~\cite{heiliger2008,henk2006}, which decouples
the middle region from the leads (see Fig.~\ref{geometry}). In this case, the decoupled system is described by
$H_{dc}=H_c+V^{dc}$. This potential difference connects the two GFs via the Dyson equation
\begin{eqnarray}
G_c 
= \tilde{G}_{dc} &-& \tilde{G}_{dc}\; ( V^{dc}_L + V^{dc}_R )\; G_c \nonumber\\
= \tilde{G}_{dc} &-& \tilde{G}_{dc}\; ( V^{dc}_L + V^{dc}_R )\; \tilde{G}_{dc} \nonumber\\
&+& \tilde{G}_{dc}\; ( V^{dc}_L + V^{dc}_R )\; \tilde{G}_{dc}\; ( V^{dc}_L + V^{dc}_R )\; G_c ,
\end{eqnarray}
where the second equation is derived by inserting the Dyson equation in itself.
Some of the arising terms can be neglected if the following assumptions are made (written schematically)
\begin{equation}
\tilde{G}_{dc}\; ( V^{dc}_L + V^{dc}_R )\; \tilde{G}_{dc}
\ll \tilde{G}_{dc}\; ( V^{dc}_L + V^{dc}_R )\; \tilde{G}_{dc}\; ( V^{dc}_L + V^{dc}_R )\; G_c
\end{equation}
and
\begin{equation}
\tilde{G}_{dc}\; ( V^{dc}_L\; \tilde{G}_{dc}\; V^{dc}_R + V^{dc}_R\; \tilde{G}_{dc}\; V^{dc}_L )\; G_c 
\ll \tilde{G}_{dc}\; ( V^{dc}_L\; \tilde{G}_{dc}\; V^{dc}_L + V^{dc}_R\; \tilde{G}_{dc}\; V^{dc}_R )\; G_c .
\end{equation}
The first assumption is obviously fulfilled if $ 1 \ll ( V^{dc}_L + V^{dc}_R )\; G_c $, which can be assured by choosing
an appropriately high decoupling potential $V^{dc}$. The second assumption makes sure that the self-energy of the leads
can be written as a sum of right and left lead self-energy. This is fulfilled if the leads are well separated because the
elements of $\tilde{G}_{dc}$ relating the left and the right leads decay exponentially with respect to the thickness of
the decoupling potential.
By comparing the remaining terms with Eq.~(\ref{SE2})
\begin{eqnarray}
G_{c}= \tilde{G}_{dc} &+& \tilde{G}_{dc}\; \Sigma_L\; G_c + \tilde{G}_{dc}\; \Sigma_R\; G_c \nonumber\\
= \tilde{G}_{dc} &+& \tilde{G}_{dc}\; V^{dc}_L\; \tilde{G}_{dc}\; V^{dc}_L\; G_c \nonumber\\
&+& \tilde{G}_{dc}\; V^{dc}_R\; \tilde{G}_{dc}\; V^{dc}_R G_c
\end{eqnarray}
one can identify the self-energies in the KKR scheme with
\begin{eqnarray}
\Sigma_L &=& V^{dc}_L\; \tilde{G}_{dc}\; V^{dc}_L \nonumber,\\
\Sigma_R &=& V^{dc}_R\; \tilde{G}_{dc}\; V^{dc}_R \label{SEKKR}.
\end{eqnarray}

\subsection{NEGF in the Korringa-Kohn-Rostoker basis}
In the KKR the Green's function (GF) in cell centered coordinates can be written
as~\cite{zabloudil}
\begin{eqnarray}\label{GFKKR}
G(\vek r + \vek R_n, \vek r'+ \vek R_{n'};z) 
&=& \delta_{nn'}\; g^n_{sc}(\vek r,\vek r';z) \nonumber\\
&+& R^n(\vek r;z)\; g^{nn'}(z)\; R^{n'}(\vek r';z)^\times ,
\end{eqnarray}
where $R^n$ is the regular solution of the single scatterer (isolated atom) at site $n$, $g^n_{sc}$ is the single
scatterer GF, and $g^{nn'}(z)$ is the so called structural GF. We use the vector notation of Ref.~\cite{zabloudil} where
$(\cdot)^\times$ conjugates only the angular functions but not the radial part and transposes all vectors. All these
quantities are vectors/matrices in the angular momentum space, which is $L=(l,m,s)$ for non-relativistic treatment or the
$Q=(\kappa,\mu)$ representation in the full-relativistic case.

By inserting the KKR GF in the transport formula~(\ref{TE}) and in the density formula~(\ref{nequ_den}) one obtains
\begin{equation}\label{TEKKR}
 T(E) = \lim_{z \rightarrow E}  \mathrm{Tr} \Big[ g(z)\; \gamma_L(z,z^*)\; g(z^*)\; \gamma_R(z^*,z)\Big] ,
\end{equation}
\begin{equation}\label{denKKR}
 n^n_{L/R}(\vek r,E) = \frac{i}{\pi} \lim_{z \rightarrow E} R^n(\vek r;z)\; g^{<,n}_{L/R}(z)\; R^{n}(\vek
 r;z^*)^\times ,
\end{equation}
with
\begin{equation}\label{denKKR1}
 g^{<,n}_{L/R}(z) = \sum_{kl} g^{nk}(z)\; \gamma_{L/R}^{kl}(z,z^*)\; g^{ln}(z^*),
\end{equation}
where the trace runs over atoms and basis representation and the quantity $\gamma_{L/R}$ is defined as
\begin{eqnarray}\
\gamma_{L/R}^{nn'}(z,z^*)&=& \int \mathrm{d}\vek r \int \mathrm{d}\vek r' \; R^n(\vek
r;z)^\times  \; \Gamma_{L/R}^{nn'}(\vek r,\vek r';z) \; R^{n'}(\vek r';z^*) \nonumber \\
\gamma_{L/R}^{nn'}(z^*,z)&=& \int \mathrm{d}\vek r \int \mathrm{d}\vek r' \; R^n(\vek
r;z^*)^\times \; \Gamma_{L/R}^{nn'}(\vek r,\vek r';z) \; R^{n'}(\vek r';z) \label{gamma0}.
\end{eqnarray}
Please note that the single scatterer GF $g^n_{sc}$ in Eq.~(\ref{GFKKR}) does not contribute in (\ref{TEKKR}) and
(\ref{denKKR1}) because only non-site-diagonal elements of the GF enter these formulas. On the other hand, the single
scatterer GF enters in Eq.~(\ref{gamma0}) as we see in the following.

Using the definition of $\Gamma(z) = i\;(\Sigma(z)-\Sigma(z^*))$ one gets
\begin{eqnarray*}
-i\; \gamma_{L/R}^{nn'}(z,z^*) &=& \int \mathrm{d}\vek r \int \mathrm{d}\vek r'\; R^n(\vek r;z)^\times  \; \left(
\Sigma_{L/R}^{nn'}(\vek r,\vek r';z)-\Sigma_{L/R}^{nn'}(\vek r,\vek r';z^*) \right) \; R^{n'}(\vek r';z^*) .
\end{eqnarray*}
With the use of the self-energy Eq.~(\ref{SEKKR}) and the KKR GF Eq.~(\ref{GFKKR}) $\gamma_{L/R}$ is given by
\begin{eqnarray*}
-i\; \gamma_{L/R}^{nn'}(z,z^*)
&=& \int \mathrm{d}\vek r \int \mathrm{d}\vek r' \; R^n(\vek r;z)^\times \; \Big[ \\
& & V^n_{dc}(\vek r) \; \left( \; \tilde{g}^n_{dc,sc}(\vek r,\vek r';z)\; \delta_{nn'} +
R^n_{dc}(\vek r;z) \; \tilde{g}^{nn'}_{dc}(z) \; R^{n'}_{dc}(\vek r';z)^\times 
\right ) \; V^{n'}_{dc}(\vek r') \\
&-& V^n_{dc}(\vek r) \; \left( \tilde{g}^{n}_{dc,sc}(\vek r,\vek r';z^*)\; \delta_{nn'} +
R^{n}_{dc}(\vek r;z^*) \; \tilde{g}^{nn'}_{dc}(z^*) \; 
R^{n'}_{dc}(\vek r';z^*)^\times \right ) \; V^{n'}_{dc}(\vek r') \\
& & \Big] \; R^{n'}(\vek r';z^*) \\
&=& \int \mathrm{d}\vek r \; R^n(\vek r;z)^\times \; V^n_{dc}(\vek r) \; \int \mathrm{d}\vek r' \; 
\tilde{g}^n_{dc,sc}(\vek r,\vek r';z) \; V^{n'}_{dc}(\vek r') \; 
R^{n'}(\vek r';z^*)\; \delta_{nn'} \\
&-& \int \mathrm{d}\vek r \; R^n(\vek r;z)^\times \; V^n_{dc}(\vek r) \int \mathrm{d}\vek r' \; 
\tilde{g}^{n}_{dc,sc}(\vek r,\vek r';z^*) \; V^{n'}_{dc}(\vek r') \; 
R^{n'}(\vek r';z^*)\; \delta_{nn'}  \\
&+& \int \mathrm{d}\vek r \; R^n(\vek r;z)^\times \; V^n_{dc}(\vek r) \;
 R^n_{dc}(\vek r;z) \; \tilde{g}^{nn'}_{dc}(z) \; \int \mathrm{d}\vek r' \; R^{n'}_{dc}(\vek
r';z)^\times \; V^{n'}_{dc}(\vek r') \; R^{n'}(\vek r';z^*) \\
&-& \int \mathrm{d}\vek r \; R^n(\vek r;z)^\times \; V^n_{dc}(\vek r) \; 
R^{n}_{dc}(\vek r;z^*) \; \tilde{g}^{nn'}_{dc}(z^*) \; \int \mathrm{d}\vek r' \; 
R^{n'}_{dc}(\vek r';z^*)^\times \; V^{n'}_{dc}(\vek r') \; R^{n'}(\vek r';z^*),
\end{eqnarray*}
where for $\gamma_L$ ($\gamma_R$) both site indices $n$, $n'$ are restricted to sites where $V_{dc,L}$ ($V_{dc,R}$)
is nonzero. The Lippmann-Schwinger equation for the single scatterer solution
\begin{equation}
\int \mathrm{d}\vek r \; R^n(\vek r;z)^\times \; V^n_{dc}(\vek r)\; \tilde{g}^n_{dc,sc}\; (\vek r,\vek
r';z) = R^n(\vek r';z)^\times - R^n_{dc}(\vek r';z)^\times ,
\end{equation}
reduces the expression to the simple form
\begin{eqnarray}
-i\; \gamma_{L/R}^{nn'}(z,z^*)
&=& \delta_{nn'}\; \left( - \Delta\tilde t^n(z,z^*)  + \Delta\tilde t^n(z^*,z)
\right ) \nonumber \\
&+& \Delta t^n(z)\; \tilde{g}^{nn'}_{dc}(z)\; \Delta\tilde t^{n'}(z^*,z) - \Delta\tilde
t^n(z,z^*)\; \tilde{g}^{nn'}_{dc}(z^*)\; \Delta t^{n'}(z^*) ,
\label{gamma1}
\end{eqnarray}
where the definition of the t-matrix $\Delta t^n(z)$ and a to the t-matrix similar expression $\Delta\tilde t^n(z,z^*)$
were used
\begin{eqnarray}\label{deltat}
\Delta t^n(z) = \int \mathrm{d}\vek r\; R^n_{dc}(\vek r;z)^\times\; V^n_{dc}(\vek r)\; R^{n}(\vek r;z) \\
\Delta\tilde t^n(z,z^*) = \int \mathrm{d}\vek r\; R^n_{dc}(\vek r;z)^\times\; V^n_{dc}(\vek r)\; R^{n}(\vek r;z^*) .
\end{eqnarray}
In a similar procedure one can calculate the quantity $\gamma_{L/R}(z^*,z)$ with interchanged energy variables
\begin{eqnarray}
-i\; \gamma_{L/R}^{nn'}(z^*,z) &=& \delta_{nn'}\; \left( - \Delta\tilde t^{n}(z,z^*)^*  + \Delta\tilde t^{n}(z,z^*)^T
\right ) \nonumber\\ &+& \Delta\tilde t^{n}(z,z^*)^*\; \tilde{g}^{nn'}_{dc}(z)\; \Delta t^{n'}(z)^T - \Delta t^n(z)^*\;
 \tilde{g}^{nn'}_{dc}(z^*)\; \Delta\tilde t^{n'}(z,z^*)^T ,
\label{gamma2}
\end{eqnarray}
where $(\cdot)^T$ stands for the transposed matrix.

This is the representation of the operator $\Gamma$ in the KKR basis similar to the structural GF $g^{nn'}(z)$. The task
of calculating the self-energies is transformed into the task of calculating the structural GF of the system with the
decoupling potential. However, in the screened KKR~\cite{zabloudil} one solves a system with a repulsive potential, the
reference system, which already fulfills the requirements of the decoupling potential. With the help of the screened KKR
and the decimation technique the surface GF of a system can be calculated, which turns out to be the GF of the
decoupled system needed in Eqs.~(\ref{gamma1}),~(\ref{gamma2}).
Another point is the usage of $z$ and $z^*$ in Eqs.~(\ref{gamma1}),~(\ref{gamma2}), which requires solving the involved
terms for two energies. One can make use of the following properties
\begin{eqnarray}
\tilde{g}_{dc}(z) &=& \tilde{g}_{dc}(z^*)^\dagger \\
\Delta t^n(z) &=& \Delta t^n(z^*)^\dagger \\
\Delta\tilde t^n(z,z^*) &=& \Delta\tilde t^n(z^*,z)^\dagger ,
\end{eqnarray}
to decrease the numerical effort to solve the t-matrix $t^n$ and the structural GF $\tilde{g}_{dc}$ only for the energy
$z$. However, the modified t-matrix $\Delta\tilde t^n$ given by Eq.~(\ref{deltat}) still requires the single scattering
wave functions for $z$ and $z^*$. In the non-relativistic case with the use of the atomic sphere approximation the
scattering solutions $R^{n}(\vek r;z)$ obtain a simple phase factor from changing the energy from $z$ to
$z^*$~\cite{achilles2013pre}. The screened reference system in the KKR is usually calculated for this case and is then
transformed in the needed representation (full-relativistic or full-potential or both). This is exact and no
approximation, because the reference system has no physical meaning. Therefore, one can simply calculate $R^n_{dc}(\vek
r;z^*)$ from $R^n_{dc}(\vek r;z)$ even for the full-potential and full-relativistic case. With this one can obtain
$\gamma$ by solving all quantities only at energy $z$.

\subsection{Coherent Potential Approximation}
In the single site KKR-CPA the effective medium corresponds to placing (site-diagonal) effective medium t-matrices
$\overline{t}^{nn'}=\delta_{nn'}\; \overline{t}^n$ on all alloy sites. The $\overline{t}^n$ are chosen to restore the
periodicity of the underlying lattice and hence the effective medium Green's function (GF) defined by:
\begin{equation}\label{equ_G_c}
\overline{g}(z)=\left[ \stackrel{o}{g}(z)^{-1} - \overline{t}(z) \right]^{-1},
\end{equation}
can be calculated using standard methods (lattice Fourier transform, decimation~\cite{zabloudil}), where
$\stackrel{o}{g}$ is the free GF. Following Butler~\cite{butler1985}, we can show that the alloy GF for a fixed
configuration and the effective medium GF are related by
\begin{equation}\label{equ_tau_tau_c}
g^{nn'}(z)=\overline{g}^{nn'}(z) + \sum_{n''\; n'''} \overline{g}^{nn''}(z)\; T^{n'' n'''}(z)\;
\overline{g}^{n'''n'}(z) .
\end{equation}
The alloy T-matrix can be calculated from
\begin{eqnarray}
T^{nn'}(z)&=&x^{n}(z)\left( \delta_{nn'} + \sum_{n'' \neq n} 
\overline{g}^{nn''}(z)\; T^{n'' n'}(z)\right) \nonumber \\
&=&\left( \delta_{nn'} + \sum_{n'' \neq n} T^{nn''}(z)\; \overline{g}^{n''n'}(z)
\right)x^{n'}(z)
\label{equ_T} ,
\end{eqnarray}
where $x^n$ is given by 
\begin{equation}
x^n(z)=\left[1-(t^n(z)-\overline{t}^n(z))\; \overline{g}^{nn}(z) 
\right]^{-1}(t^n(z)-\overline{t}^n(z)).
\end{equation}
Here, the lattice sites $t^n$ are occupied according to the fixed configuration. The $x^n$ describe the additional
scattering of the alloy relative to the effective medium.
Using the single site approximation (SSA)~\cite{velicky1968} leads to the decoupling of the averaged T-Matrix $\langle
T\rangle=\langle x^n \rangle \left( \delta_{nn'} + \sum_{n'' \neq n} \overline{g}^{nn''}\; \langle T^{n''n'}\rangle
\right)$. Therefore, if the single site CPA condition $\langle x^{n} \rangle=0$ is fulfilled, we find $\langle
T\rangle=0$ and that the GF of the configurational averaged system is identical to the GF of the effective medium
$\langle g^{nn'}\rangle=\overline{g}^{nn'}$.
For the non-equilibrium applications we need to calculate alloy averages like $\langle G\; A\; G \rangle$ for some
operator $A$. In order to express this in terms of effective medium GF one demands the relation
\begin{equation}
\langle g(z) \; A(z) \; g(z^*) \rangle = \overline{g}(z) \; A(z) \; \overline{g}(z^*)
+ \overline{g}(z) \; \Omega_A(z) \; \overline{g}(z^*),
\end{equation}
where the first term on the right-hand side represents the coherent contribution while the second term defines the
non-equilibrium vertex corrections (NVC) $\Omega_A$. As we will show, they are the result of multiple-scattering
relative to coherent but damped motion in the effective medium. Thus, they can be interpreted as accounting for
diffusive contributions. Note that the NVC are specific to the operator A.
In our case the operator $A$ does not depend on the alloy configuration and using Eq.~(\ref{equ_tau_tau_c}) and SSA we
find that
\begin{equation}
\Omega_A(z) = \langle T(z) \; \overline{g}(z) \; A(z) \; \overline{g}(z^*) \; T(z^*) \rangle
\end{equation}
Using Eq.~(\ref{equ_T}) and the SSA we find that the vertex corrections are site-diagonal $\Omega_A^{n
n'}=\delta_{n n'}\; \Omega_A^{n}$ and a closed set of equations can be derived~\cite{velicky1968,carva2006,ke2008}:
\begin{eqnarray}
\Omega_A^{n}(z) 
&=& \langle x^n(z) \;  \left[ \overline{g}(z) \; A(z) \; \overline{g}(z^*) \right]_{n n} \; x^n(z^*) \rangle
\nonumber \\ &&+ \sum_{n'\neq n} \langle x^n(z) \; \overline{g}^{n n'}(z) \; \Omega_A^{n'}(z) \; \overline{g}^{n' n}(z^*) \;
x^n(z^*) \rangle \nonumber  \\ 
&=& \sum_{\alpha} c^n_\alpha  \; x^n_\alpha(z) \; \left[ \overline{g}(z) \; A(z) \; \overline{g}(z^*) \right]_{n n}
\;x^n_\alpha(z^*) \nonumber  \\ &&+ \sum_{\alpha, n'\neq n} c^n_\alpha \; x^n_\alpha(z) \; \overline{g}^{n
n'}(z) \; \Omega_A^{n'}(z) \; \overline{g}^{n'
n}(z^*) \; x^n_\alpha(z^*)
\label{equ_nvc} ,
\end{eqnarray}
where $\alpha$ enumerates the species on site $n$, $c^n_\alpha$ is the particular concentration, and $x^n_\alpha$ is
$x^n$ when the site is occupied by $\alpha$. The NVC are zero for non-alloy sites, thus, one has to calculate the
NVC only for CPA sites.

In order to make use of the two-dimensional translational symmetry, we complete the sum over $n'$ and transform the
infinite lattice sums to $\vek k_\parallel$-space integrations over the first Brillouin zone. Writing 
$n \rightarrow \vek S+ \vek T_\parallel$, where $\vek T_\parallel$ is a two-dimensional lattice vector and $\vek S$
belongs to the primitive cell, we find
\begin{eqnarray}
\Omega_A^{\vek S}(z) &=& \sum_{\alpha} c^{\vek S}_\alpha  \; x^{\vek S}_\alpha(z) \;  \left[\sum_{\vek
k_\parallel} \; \overline{g}(\vek k_\parallel; z)  \;A(\vek k_\parallel;z) \;
\overline{g}(\vek k_\parallel; z^*) \right]_{\vek S \vek S} x^{\vek S}_\alpha(z^*) \nonumber \\ 
 &&- \sum_{\alpha} c^{\vek S}_\alpha \; x^{\vek S}_\alpha(z) \; \overline{g}^{\vek S \vek S}(z) \; 
\Omega_A^{\vek S}(z) \; \overline{g}^{\vek S \vek S}(z^*)\;  x^{\vek S}_\alpha(z^*)
\nonumber \\ 
 &&+ \sum_{\alpha} c^{\vek S}_\alpha \; x^{\vek S}_\alpha(z) \left[ \sum_{\vek k_\parallel} 
\; \overline{g}(\vek k_\parallel; z) \; \Omega_A(z) \; \overline{g}(\vek k_\parallel; z^*) \right]_{\vek S \vek S}
x^{\vek S}_\alpha(z^*) .
\label{equ_nvc1}
\end{eqnarray}
Note that $\Omega_A$ is site-diagonal and hence does not depend on $\vek k_\parallel$. Therefore, we could pull
$\Omega_A$ out of the $k_\parallel$-sum and solve the linear equation by matrix inversion. Alternatively, one can obtain
$\Omega_A$ directly by iterating Eq.~(\ref{equ_nvc1}). The matrix inversion is in general faster than the iterative
procedure, but needs the full coupling matrix, which can become very large. The iterative procedure has the advantage
that it allows for a splitting the problem in smaller tasks (the vertex correction for one atom), which can be utilized
in parallel computing. Additionally, only a part of the complete matrix is needed for every single task, which results in
a smaller memory consumption per task. It should be noted that the inversion has to be done only ones after the
Brillouin zone integration.
The theory of NVC can be directly applied to the transport equation. By noting that $\gamma_{L/R}$ does not depend on the
alloy configuration, we find
\begin{equation}\label{TEKKRnvc}
 \langle T(E) \rangle = \lim_{z \rightarrow E}  \mathrm{Tr} \Big[ 
 \overline{g}(z)\; \gamma_L(z,z^*)\; \overline{g}(z^*)\; \gamma_R(z^*,z)
+  \overline{g}(z)\; \Omega_{\gamma_L}(z)\; \overline{g}(z^*)\; \gamma_R(z^*,z) \Big].
\end{equation}

The averaged density at site $n$ is expressed as the weighted sum over the densities of the components
\begin{equation}\label{denpro1}
 \langle n^n(\vek r; E) \rangle = \lim_{z \rightarrow E} \sum_\alpha c^n_\alpha\; n^n_\alpha(\vek r; z) .
\end{equation}
This is necessary, since the effective medium does not provide a scattering solution $\overline{R}^n$ or single scatterer
GF $\overline{g}^n_{sc}$. The component densities can be calculated from restricted averages, where only the atom on site
$n$ is fixed. In equilibrium, they are obtained from projections of the effective medium GF~\cite{zabloudil} via the
impurity matrix $D_\alpha^n$
\begin{eqnarray}
 n^n_\alpha(\vek r; z) &=&  \langle n^n(\vek r; z) \rangle_{(n=\alpha)} = -\frac{1}{\pi} \mathrm{Im} \left[
 R_\alpha^{n}(\vek r;z) \; D^n_\alpha(z) \;  \overline{g}^{nn}(z) \; R_\alpha^{n}(\vek
r;z)^\times \right] ,\\
D_\alpha^n(z) &=& \left[ 1 - ( t^n_{\alpha}(z) - \overline{t}^n(z) ) \;\overline{g}^{nn}(z)
\right]^{-1}
\end{eqnarray}

For the non-equilibrium density in the presence of CPA alloys, $g^{<,n}(z)$ in Eq.~(\ref{denKKR}) is
replaced by
\begin{equation}\label{denKKR2}
 \overline{g}^{<,n}(z) = \sum_{kl} \overline{g}^{nk}(z)\left( \gamma^{kl}(z,z^*)+
 \delta_{kl} \; \Omega_{\gamma}^{k}(z)\right) \overline{g}^{ln}(z^*)
\end{equation}
for all sites. This means that the NVC influence the non-equilibrium density for non alloy sites also.

Again, on alloy sites component densities must be used. Following Ref.~\cite{ke2008}, the component non-equilibrium
densities, for the special case of a binary alloy with the components $A$ and $B$, can be calculated via
\begin{equation}\label{denKKR3}
 n^n_\alpha(\vek r;z) = -\frac{i}{\pi} R_\alpha^{n}(\vek r;z) \; \overline{g}_\alpha^{<,n}(z) \; R_\alpha^{n}(\vek
r;z)^\times ,
\end{equation}
where the quantity $g_\alpha^{<,n}(z)$ is given by
\begin{eqnarray}
 \begin{aligned}
 \overline{g}_{A}^{<,n}(z) = ( t^n_{B}(z) - t^n_{A}(z) )^{-1} \left[ ( t^n_{B}(z) - \overline{t}^n(z) )\;
 \overline{g}^{<,n}(z) \right. \\ \left.  -\; \Omega_{\gamma}^n(z)\; \overline{g}^{nn}(z^*)  \right]
 /c^n_{A} \end{aligned} \\
\begin{aligned}
 \overline{g}_{B}^{<,n}(z) = ( t^n_{A}(z) - t^n_{B}(z) )^{-1} \left[ ( t^n_{A}(z) - \overline{t}^n(z) )\;
 \overline{g}^{<,n}(z) \right. \\ \left. -\; \Omega_{\gamma}^n(z)\; \overline{g}^{nn}(z^*)  \right]
 /c^n_{B} .
\end{aligned}\label{denpro2}
\end{eqnarray}
Unfortunately, the derivation in Ref.~\cite{ke2008} relies on a system of equations, which is only closed and solvable
for binary alloys. It does not allow for an obvious generalization to multicomponent alloys. Such a generalization is an
important task for future developments.

\subsection{Results}
In order to have a convincing test of the NVC we calculate the transport through an alloyed layer comparing results from
CPA calculations with and without NVC with supercell calculations. We choose a system of 10 atomic layers of FeCo alloy
in the structure of Fe connected to leads, which consist of Cu in the Fe bcc structure. For the supercells we use 16
atoms in-plane and average the results over 10 random configurations for each concentration. The self-consistent
potentials were obtained using our all-electron screened KKR code within the local spin-density approximation (LSDA) and
with a non-relativistic treatment. We use the parametrization of Vosko-Wilk-Nusair~\cite{vosko1980} for the local
exchange-correlation potentials. Since we are interested in a test of the transport properties, we use CPA potentials for
the supercell. The alloy and the Cu leads are considered in bcc-structure with $a=0.287 nm$. The calculations were
performed for concentrations in the range from $0$ to $100\%$. The integrations in the 2D BZ were performed on a uniform
mesh of $24^2$ k-points for the self-consistent calculation and $160^2$ ($40^2$) k-points at $\eta=2\cdot 10^{-5} \
\mathrm{Ry}$ for the CPA (supercell) transport calculations.

\begin{figure}
\centering
\includegraphics[width=0.45\textwidth]{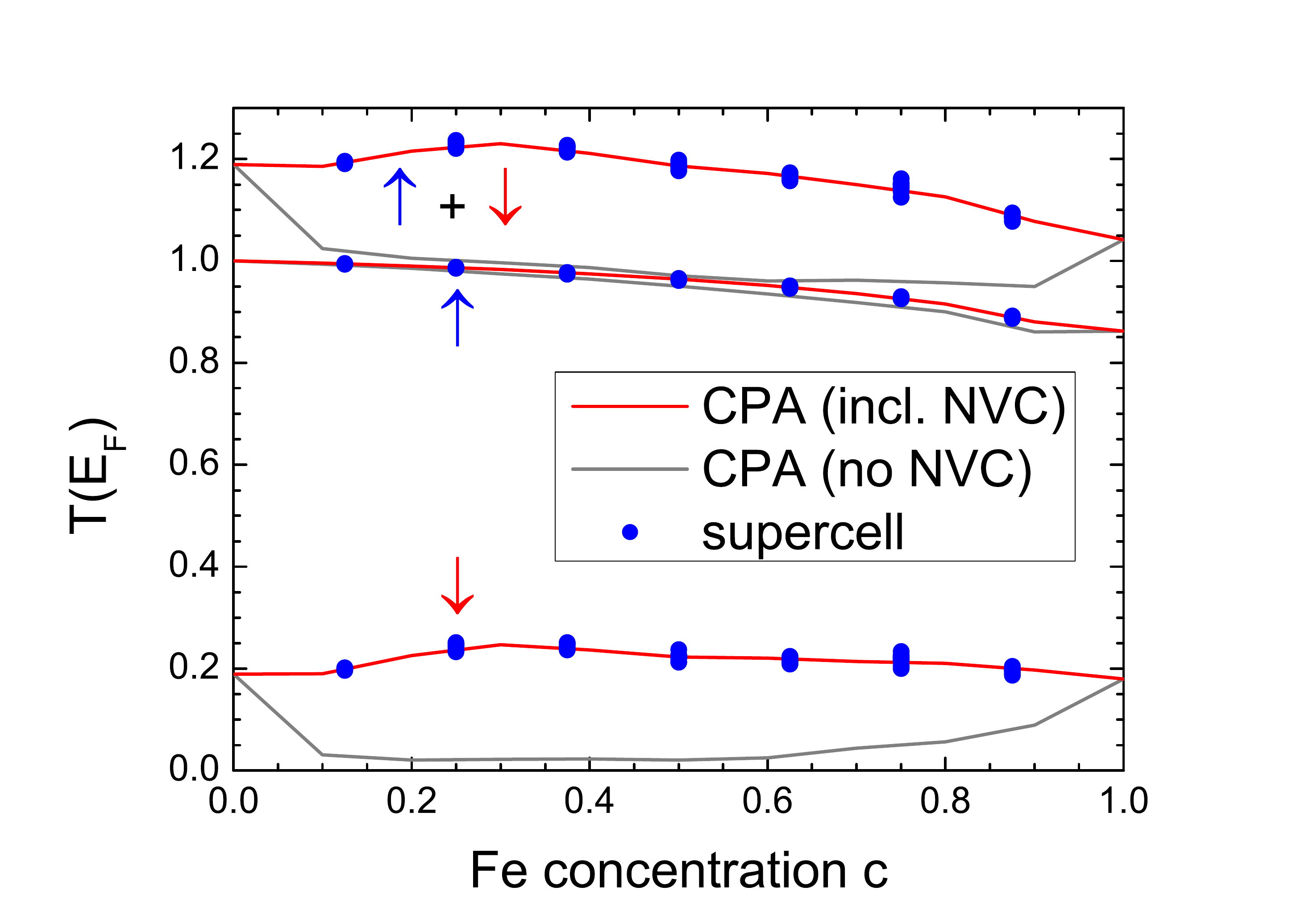}
\caption{Transmission $T(E_F)$ at the Fermi-energy through a thin Fe$_c$Co$_{1-c}$ layer, calculated with a supercell
method (dots) and the CPA (red line), also showing the results without NVC (grey line).}
\label{img_trans}
\end{figure}

\begin{figure}
\centering
\includegraphics[width=0.45\textwidth]{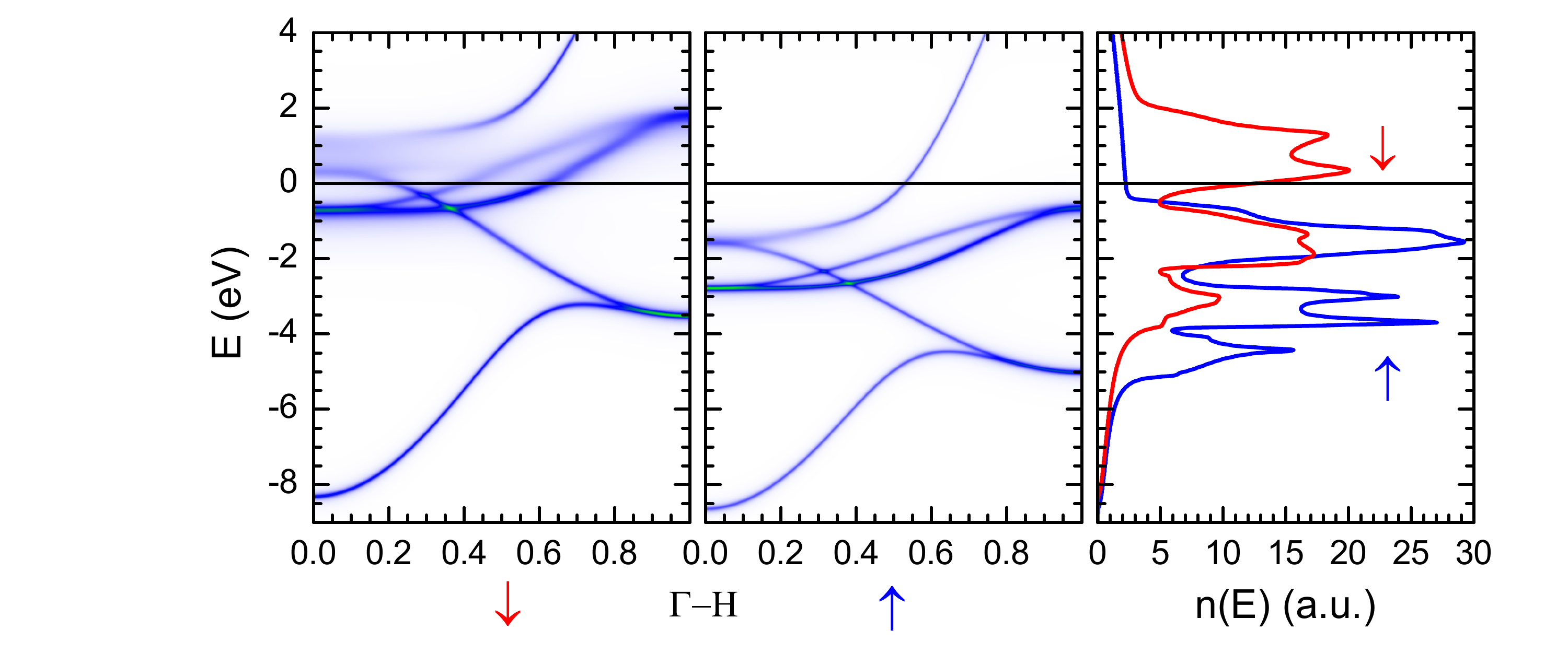}
\caption{Bloch spectral density $A^B(\vek k_\parallel,E)$ along the $\Delta$-line ($\Gamma$-H) and density of states
$n(E)$ in Fe$_{0.5}$Co$_{0.5}$ for both spin directions}
\label{img_bsfe}
\end{figure}

The results in Fig.~\ref{img_trans} show that the CPA results including NVC are in excellent agreement with the supercell
calculations and that the vertex corrections are necessary to obtain correct results. It is salient that the NVC are
dominant in the spin down channel while they are small in the spin up channel. This can be easily understood by looking
at the Bloch spectral density~\cite{faulkner1980} $A^B(\vek k_\parallel,E)$, which is shown in Fig.~\ref{img_bsfe} for
bulk Fe$_{0.5}$Co$_{0.5}$. We can see that the bands for spin down show a strong broadening at the Fermi-energy
corresponding to a strong alloy scattering. This leads to a strong effect of the vertex corrections and make them
non-negligible. However, the Bloch spectral density at the Fermi-energy of the spin up bands are rather sharp and
indicate a weak alloy scattering at this energy, which results in small vertex corrections for the spin up conductance.
This demonstrates a severe difficulty for judging the importance of the NVC, since the impact of alloy scattering can be
strongly energy dependent. Generally, neglecting the NVC (i.e.\ neglecting the diffusive current) breaks the current
conservation~\cite{velicky1968} and can lead to unphysical results.

\begin{figure}
\centering
\includegraphics[width=0.45\textwidth]{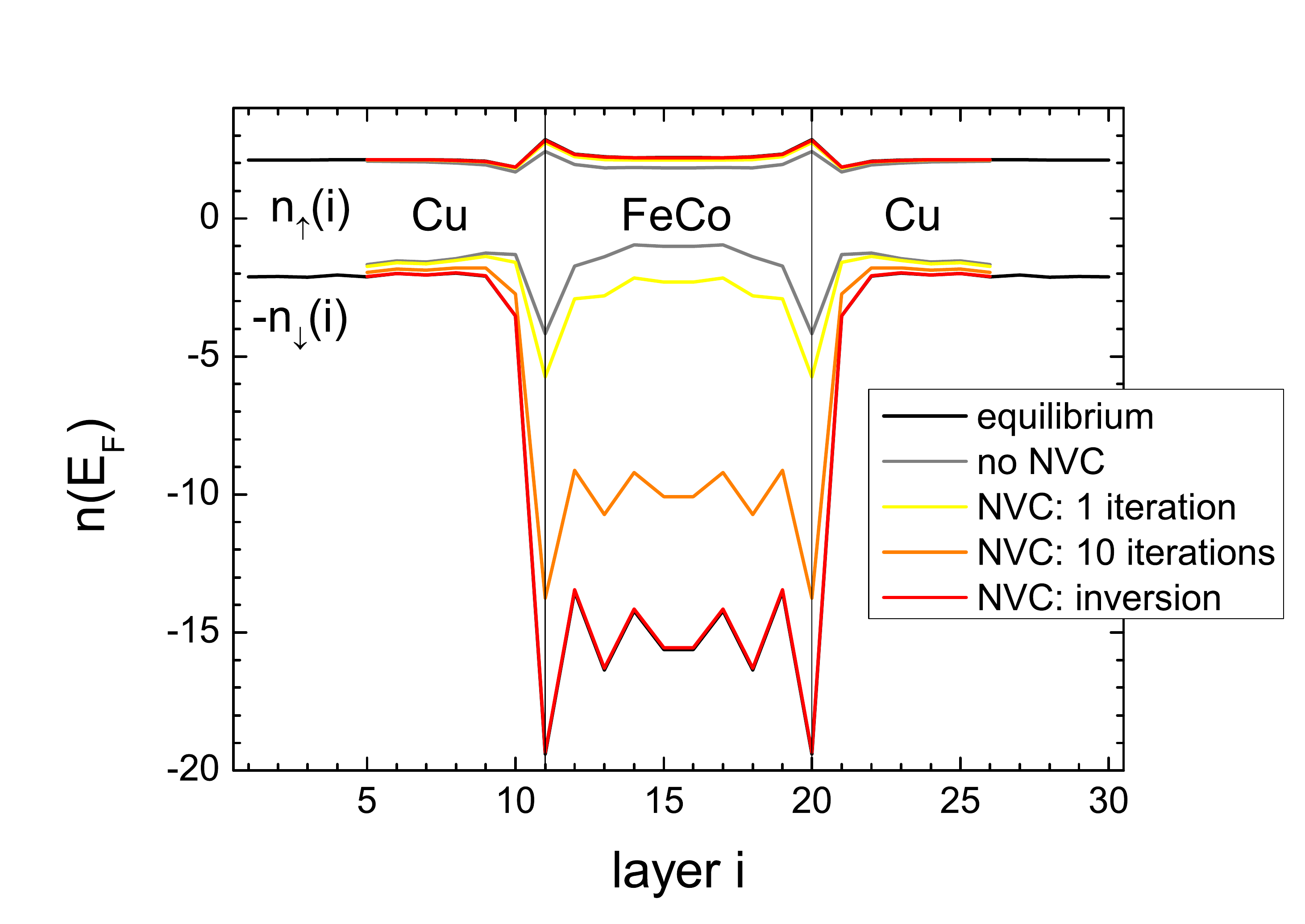}
\caption{Layer and spin resolved density of states at the Fermi-energy $n(E_f)$ in a thin layer of Fe$_{0.3}$Co$_{0.7}$
between Cu leads, comparing the equilibrium density with several stages of the NVC calculation, note that the
non-equilibrium density is not defined in the outermost layers of the leads.}
\label{img_denef}
\end{figure}

We can also use the non-equilibrium density (\ref{nequ_den}) to get an even more stringent test using the fact that
\begin{equation}
n^L+n^R=n^{equ},
\end{equation}
where $n^{equ}$ is the equilibrium density. The NVC for the non-equilibrium density are calculated using
Eq.~(\ref{equ_nvc}). It is also necessary to perform projections onto the species resolved densities (see
Eqs.~(\ref{denpro1})-(\ref{denpro2})). We perform this test for the FeCo system. The results are summarized in
Fig.~\ref{img_denef}. One can see that the test is satisfied. The iterative solution shows a sufficient convergence after
about 100 iterations. On the other side the calculation of the NVC by inversion shows a perfect matching with the
equilibrium density. We find the same agreement for the component densities (not shown).

\subsection{Conclusion}
We have implemented the coherent potential approximation (CPA) and the necessary non-equilibrium vertex corrections (NVC)
in a KKR method. This makes accurate \textit{ab initio} description of alloys in equilibrium and non-equilibrium
systems possible. The CPA includes the incoherent scattering of Bloch waves in the description. This is visible as a broadening
of the energy levels of the states, visible in the Bloch spectral density. It also leads to diffusive transport described
by the NVC. We validate our implementation by calculating transport through FeCo alloys and comparing the CPA to
supercell results. Additionally, we check an identity for the non-equilibrium density, which demonstrates the
correctness of the non-equilibrium density and the projections. Our results emphasize the importance of the NVC.

\subsection{Acknowledgements}
We acknowledge the funding provided by the German Research Foundation grant \mbox{HE 5911/1-1}.
\bibliography{bib}

\end{document}